\documentclass[aps,pra,twocolumn,showpacs,letterpaper,superscriptaddress]{revtex4-1}

\usepackage[colorlinks=true, citecolor=blue, linkcolor=blue, urlcolor=blue]{hyperref}
\usepackage{graphicx,dcolumn,longtable,epsfig}
\usepackage[usenames]{color}
\usepackage{amssymb}
\usepackage{amsmath}
\usepackage{bm}
\usepackage{footnote}
\usepackage{float}
\usepackage{subfigure}
\usepackage{color}       
\usepackage{ulem}
\usepackage[T1]{fontenc}

\newcommand{\be}{\begin{equation}}
\newcommand{\ee}{\end{equation}}

\def\bea{\begin{eqnarray}}
\def\eea{\end{eqnarray}}

\begin{document}

\title{Thermocrystallization of lattice dipolar bosons coupled to a high-finesse cavity}

\author{Yaghmorassene Hebib }
\affiliation{Department of Physics, Clark University, Worcester, Massachusetts 01610, USA}
\author{Chao Zhang}
\email{zhangchao1986sdu@gmail.com}
\affiliation{Department of Modern Physics, University of Science and Technology of China, Hefei, Anhui 230026, China}
\affiliation{Hefei National Laboratory, University of Science and Technology of China, Hefei, Anhui 230088, China}
\author{Massimo Boninsegni}
\affiliation{Department of Physics, 
University of Alberta,  
Edmonton,  Alberta,  Canada,  T6G 2H5 
}
\author{Barbara Capogrosso-Sansone}
\affiliation{Department of Physics, Clark University, Worcester, Massachusetts 01610, USA}

\begin{abstract}
Investigating finite temperature effects on quantum phases is key to their experimental realization. Finite temperature, and the interplay between quantum and thermal fluctuations can undermine properties and/or key features of quantum systems but they can also bring upon interesting phenomena. In this paper, we present a comprehensive investigation of the finite temperature phase diagram of two-dimensional lattice dipolar bosons coupled to a high-finesse optical cavity. Interestingly, we observe that checkerboard density-density correlations are enhanced at finite temperature. Indeed, we found that finite temperature drives a superfluid ground state into a normal state which will then develop checkerboard order at higher temperatures. We show that this effect is solely due to the cavity-mediated interactions. We also confirm that the supersolid checkerboard phase survives for a wide range of filling factors up to temperature scale of the order of half hopping amplitude, while the checkerboard diagonal order can survive up to temperatures of a few hopping amplitudes.   
\end{abstract}

\maketitle

\section{Introduction}

The experimental realization of long-range interactions with ultracold gases~\cite{chomaz_2023,polarMol,RevModPhys.82.2313,Low_2012,EsslingerNature,Farokh:2021,ritsch_cold_2013, aspelmeyer_cavity_2014} has opened up avenues for exploring intriguing  quantum phases, novel collective behaviors, and unconventional many-body quantum states that were previously inaccessible. This has sparked a  burgeoning interest in theoretical investigation of such systems~\cite{Trefzger_2011,RevModPhys.95.035002}.
An interesting setup was theoretically discussed for a system of lattice dipolar bosons coupled to high finesse cavities in two-dimensions~\cite{PhysRevA.107.043318}. 
This setup encompasses both dipolar interactions, which follow a $1/r^3$ decay, and cavity-mediated infinite-range interactions. In its ground state, the system realizes superfluid, checkerboard solid, checkerboard supersolid, and incompressible phases. One of the main findings of ~\cite{PhysRevA.107.043318} is that the checkerboard supersolid phase can exist across a broader range of particle densities compared to the case with no cavity-mediated interactions. These unbiased numerical findings suggest the practical feasibility of achieving both solid and supersolid phases experimentally using magnetic atoms and polar molecules coupled to a cavity. A key question for experimental realization of such phases is the role of thermal fluctuations.

Finite temperature, and the interplay between quantum and thermal fluctuations can undermine properties and/or key features of quantum systems but also bring upon interesting phenomena. In particular, the robustness of the liquid phase at low temperature is crucially underlain by quantum-mechanical exchanges, in Bose systems. Indeed, it has been shown \cite{Boninsegni2012} that even condensed $^4$He, long believed to remain a fluid down to zero temperature due to atomic zero-point motion alone, would in fact undergo thermo-crystallization at finite temperature, in the absence of quantum statistics (i.e., if $^4$He atoms were distinguishable quantum particles). A similar effect of reentrance of crystalline order at finite temperature, arising as exchanges are suppressed with the reduction of the thermal wavelength, has also been predicted for small clusters of parahydrogen, which undergo ``quantum melting'' as the temperature $T\to 0$ \cite{Mezzacapo2006,Mezzacapo2007}. Moreover, this interesting behavior has also been observed in experiments with ultracold dipolar gases. In~\cite{Sanchez-Baena:2023te}, authors observe that  density-density correlations are enhanced at finite temperature. 

In this paper, we study the finite temperature phase diagram of  dipolar bosons coupled to a high-finesse cavity and trapped in  a square lattice. By means of path-integral Monte Carlo, we perform simulations of the extended Bose-Hubbard model and determine critical temperatures for disappearance of diagonal and off-diagonal order. Interestingly, we observe that checkerboard density-density correlations are enhanced at finite temperature. Indeed, finite temperature drives a superfluid ground state into a normal state which will then develop checkerboard order at higher temperatures.
This paper is organized as follows: In Sec.~\ref{sec:sec2}, we introduce the Hamiltonian of the system. In Sec.~\ref{sec:sec3}, we discuss various phases and the corresponding order parameters. In Sec.~\ref{sec:sec4}, we present the finite temperature phase diagrams of the system for fixed values of the onsite and cavity-mediated interactions; we outline our conclusions in Sec.~\ref{sec:sec5}.

\section{Hamiltonian}
\label{sec:sec2}
We consider a system of dipolar bosons trapped in a square optical lattice and coupled to a high-finesse optical cavity. The dipoles are aligned perpendicular to the optical lattice plane, ensuring that the dipolar interaction is purely repulsive and isotropic. Within the single-band approximation, this system is governed by the following extended Bose-Hubbard Hamiltonian~\cite{EsslingerNature,Dogra:2016hy}:

\begin{equation}\label{eq:H}
\begin{aligned}
H = &-t \sum_{\langle ij \rangle} a_i^\dagger a_j + \frac{U_s}{2} \sum_i n_i(n_i-1) + \frac{V_{\text{dip}}}{2} \sum_{i,j} \frac{n_i n_j}{r_{ij}^3} \\
&- \frac{V_{\text{ca}}}{L^2}\left(\sum_{i \in e}n_i - \sum_{j \in o} n_j\right)^2 - \mu \sum_i n_i.
\end{aligned}
\end{equation}

Here, the first term represents the kinetic energy, characterized by hopping amplitude $t$. The summation over $\langle ij \rangle$ signifies that the sum is taken over the nearest neighboring sites. The operators $a^{\dagger}$ and $a$ are the bosonic creation and annihilation operators, respectively, obeying the standard bosonic commutation relations. The second term in the Hamiltonian describes the on-site repulsive interaction with a strength $U_s$. In this context, $n_i = a_i^{\dagger} a_i$ denotes the number operator for the particles at site $i$. The third term is the dipolar interaction between the sites, with dipolar interaction strength $V_{\text{dip}}$ and $r_{ij} = |r_i - r_j|$ as the relative distance between sites $i$ and $j$. The fourth term is the cavity-mediated long-range interaction, characterized by the interaction strength $V_{\text{ca}}$, with the summations over $i \in e$ and $j \in o$ indicating sums over even and odd lattice sites, respectively. The final term is the chemical potential $\mu$.

\section{Method and Order parameters}
\label{sec:sec3}
The results we will present in the following are based on path integral quantum Monte Carlo by the worm algorithm~\cite{PROKOFEV1998253}.
We consider a square lattice of sizes $L = 16, 20, 24, 30$ (we set our unit of length to be the lattice step, $a=1$). We impose periodic boundary conditions in both spatial dimensions.
To properly account for the long-range dipolar interaction, Ewald summation was utilized. The inverse temperature is denoted by $\beta = 1/T$ (in our units, the Boltzmann constant $k_B=1$).

To characterize the various phases, we calculated superfluid density $\rho_s$ and structure factor $S(\pi,\pi)$.
The superfluid density is calculated in terms of winding numbers~\cite{Ceperley:1989hb}:
\begin{equation}
    \rho_s = \frac{\langle \mathbf{W^2} \rangle}{D L^{D-2} \beta},
\end{equation}
where $\langle \mathbf{W}^2 \rangle = \langle\sum_{i=1}^{D}{W_i}^2 \rangle $ is the expectation value of winding number square, $D$ is the dimension of the system (here $D=2$), $L$ is the linear system size, and $\beta$ is the inverse temperature.
The structure factor characterizes diagonal long-range order and is defined as: 
\begin{equation}
   S(\mathbf{k}) = \frac{\sum_{\mathbf{r},\mathbf{r'}} \exp[i \mathbf{k} (\mathbf{r}-\mathbf{r'})\langle n_r n_{r'} \rangle]}{N}, 
\end{equation}
with $N$ the particle number. $\mathbf{k}$ is the reciprocal lattice vector. Here, $\mathbf{k}=(\pi,\pi)$ to identify a checkerboard density pattern.

\section{Results and discussion}
\label{sec:sec4}
In the following, we work at filling factor $n=N/L^2< 1$, and fixed values of $U_s/t=20.0$ and $V_{\text{ca}}/t=2.0$. Finite cavity-mediated interaction favors a density-wave between even and odd sites.  At $T=0$ and for $V_{\rm dip}\gtrsim 1.13t$, model~(\ref{eq:H}) features a  checkerboard (CB) solid phase at $n=0.5$. Upon doping the CB solid with particles or holes, the system enters a CB supersolid phase (CBSS). For large enough doping, density-density correlations are eventually destroyed via a second-order phase transition belonging to the $(2+1)$ Ising universality class, leaving the system in a superfluid (SF) phase. For $V_{\rm dip}\lesssim 1.13t$, the system is a SF for any $n$. For $V_{\rm dip}\gtrsim 10t$ other incompressible phases are stabilized. Here, we do not consider these values of dipolar interaction. For more details, see~\cite{PhysRevA.107.043318}.

\begin{figure}
    \centering
    \includegraphics[width=0.5\textwidth]{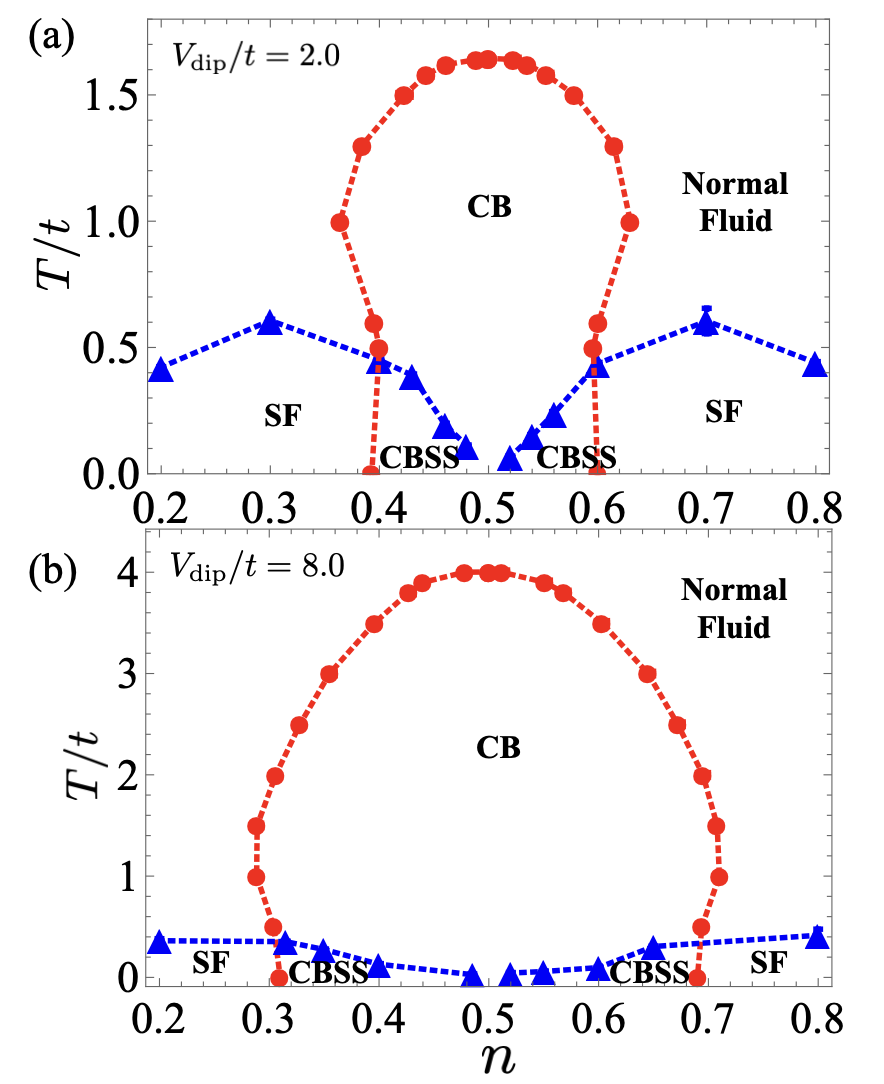} 
        \caption{(Color online) Critical temperatures for the disappearence of off-diagonal long-range order (blue triangles) and diagonal long-range order (red circles) as a function of filling factor $n$ at dipolar interaction $V_{\text{dip}}/t=2$(a) and $V_{\text{dip}}/t=8$(b).
        }
    \label{fig1}
\end{figure}

In fig.~\ref{fig1}, we show the finite $T$ phase diagram for two values of dipolar interaction $V_{\text{dip}}/t=2$(a) and $V_{\text{dip}}/t=8$(b). 
Superfluidity disappears via a Kosterlitz-Thouless (KT) type transition~\cite{Kosterlitz:1973fc} (blue triangles in fig.~\ref{fig1}), while the CB order disappears via a two-dimensional Ising-type transition (red circles). Therefore, CBSS disappears via a two-step process with two critical temperatures $T_{\rm{KT,SS}}$ and $T_{\rm{CB,SS}}$ corresponding to the disappearance of off-diagonal and diagonal long-range order respectively. Other than for $n$ values in the vicinity of the SF-CBSS transition at zero temperature, $T_{\rm{KT,SS}}<T_{\rm{CB,SS}}$ so that the off-diagonal order disappears first and the supersolid phase becomes a liquid-like phase with finite density-density correlations.

\begin{figure}
    \centering
    \includegraphics[width=0.5\textwidth]{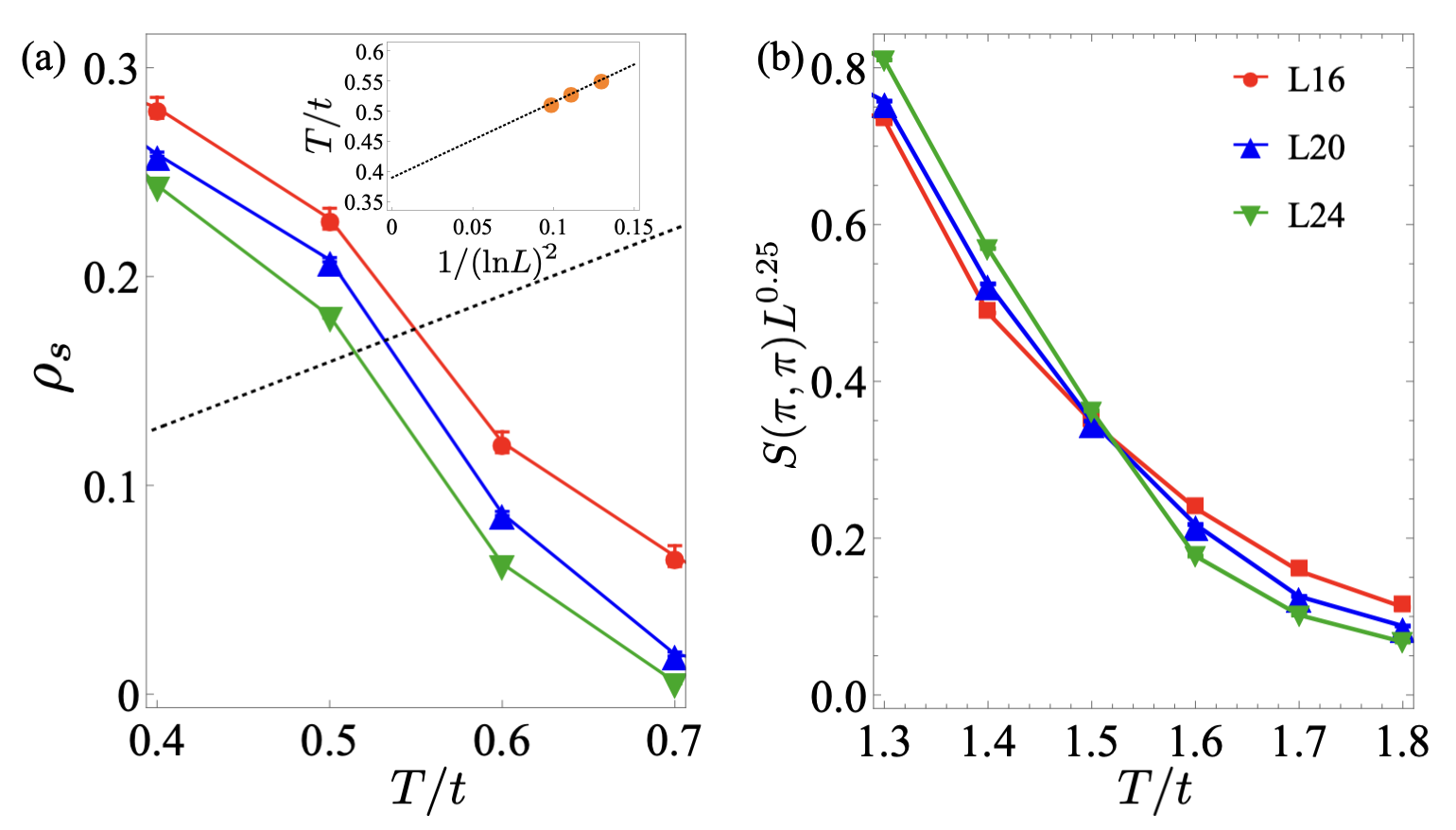}
    \caption{ $V_{\text{dip}}/t=2$ and $n=0.43$. Upon increasing the temperature, thermal fluctuations destroy the checkerboard supersolid phase in favor of a normal fluid in two steps. First, superfluidity is destroyed via a Kosterlitz–Thouless phase transition and the checkerboard supersolid becomes a liquid-like phase with finite density-density correlations. Then, the checkerboard solid order melts into a normal fluid via a two-dimensional Ising transition. In (a) we show $\rho_s$ as a function of $T/t$ for $L=16$ (red), 20 (blue), 24 (green). The dotted line is $T/t\pi$. Inset: intersection points between the $T/t\pi$ line and the $\rho_s$ versus $T/t$ curves for each $L$ are used to extract $T_c/t\sim0.39 \pm 0.01$. (b) scaled structure factor with $2\beta/\nu=0.25$ for $L=16, 20, 24$. The crossing determines the critical temperature $T_c/t = 1.52\pm0.02$.}
    \label{fig2}
\end{figure}

In fig.~\ref{fig2}, we show details on how critical temperatures are calculated. In fig.~\ref{fig2} (a), we show the procedure followed for KT-transition temperatures. The superfluid density $\rho_s$ is plotted as a function of $T/t$ for $L=16$, 20, 24 at $V_{\text{dip}}/t=2$ and $n=0.43$. In the thermodynamic limit, a universal jump is observed at the critical temperature given by $\rho_s(T_c)=2m k_B T_c/\pi \hbar^{2}$. Here, $m$ is the effective mass in the lattice, $m=\hbar^2/2ta^2$. In a finite size system this jump is smeared out as shown in Fig.~\ref{fig2}(a). To extract the critical temperature in the thermodynamic limit, we apply finite-size scaling to $T_c(L)$. From renormalization-group analysis  one finds $T_c(L)=T_c(\infty)+\frac{c}{\ln^2(L)}$, where $c$ is a constant and $T_c(L)$ is determined from $\rho_s(T_c,L)=2m k_B T_c/\pi \hbar^{2}$~\cite{PhysRevLett.39.1201,Kosterlitz_1974,Ceperley:1989hb}.
The dotted line in fig.~\ref{fig2}(a) corresponds to $\rho_s=T/t\pi$ ( $\hbar=1$, $k_B=1$, lattice step $a=1$).  Its intersection points with each $\rho_s$ vs. $T/t$ curve are used to find $T_c$ as shown in the inset. Here. we find $T_c/t=0.39 \pm 0.01$. Above this temperature the system is in a a liquid-like phase with finite density-density correlations. The solid order will eventually disappear via a two-dimensional Ising transition. In fig.~\ref{fig2}(b),  we use standard finite size scaling and plot the scaled structure factor $S(\pi,\pi)L^{2\beta/\nu}$, with $2\beta/\nu=0.25$  as a function of $T/t$ for $L=16$, 20, 24. The crossing indicates a critical temperature $T_c/t = 1.52\pm 0.02$. 

\begin{figure}
    \centering
    \includegraphics[width=0.48\textwidth]{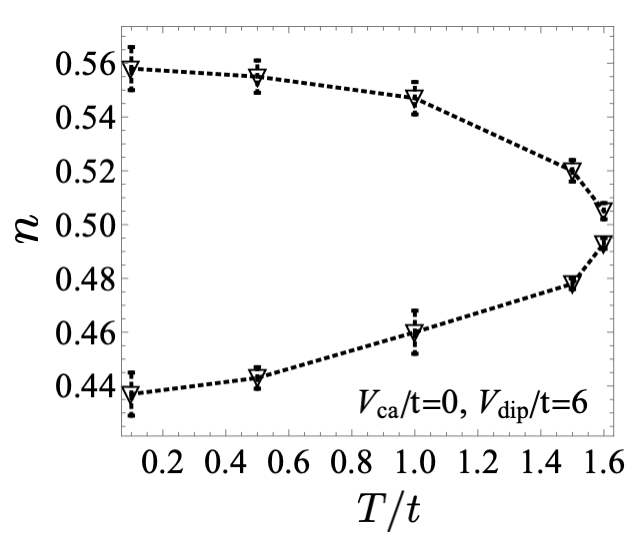}
    \caption{CB density-density correlations are finite inside the lobe and non-existent outside the lobe. In the absence of cavity-mediated interactions, no re-entrant behavior is observed. Transition points have been determined using standard finite-size scaling arguments as discussed in fig.~\ref{fig2}(b) with the difference that here we have scanned through densities at fixed T, rather than the opposite as shown in fig.~\ref{fig2}(b).}
    \label{fig3}
\end{figure}

Interestingly, the transition line for disappearance of diagonal order features a re-entrant behavior so that the system develops density-density correlations at densities for which the ground-state is SF. We notice that the onset of this re-entrant behavior is concurrent with the disappearance of  superfluidity. More specifically, in the proximity of the CBSS-SF transition at zero temperature, there exists a range of densities for which the ground state is a SF. Upon increasing temperature, the SF state disappears through a KT-transition in favor of a normal fluid, i.e. a state where quantum statistics has become less relevant. By further increasing $T$, the normal fluid develops CB-type density-density correlations. This effect is solely due to the cavity-mediated interaction. This interaction mimics an external pinning potential and discourages exchanging of indistinguishable particles. As a result, the system behaves more classical-like and supports density-density correlations~\cite{Boninsegni2012}. In fig.~\ref{fig3}, we consider the case of no cavity-mediated interaction for which the CB order exists at $V_{\text{dip}}/t\gtrsim 4.75$~\cite{PhysRevA.90.043635}.  Here, we plot the $n$ vs $T/t$ line at $V_{\text{dip}}/t=6$ for which CB density correlations disappear so that outside the lobe the CB solid order is non-existent. We do not observe any re-entrant behavior. 

\begin{figure}[ht]
    \centering
    \includegraphics[width=0.48\textwidth]{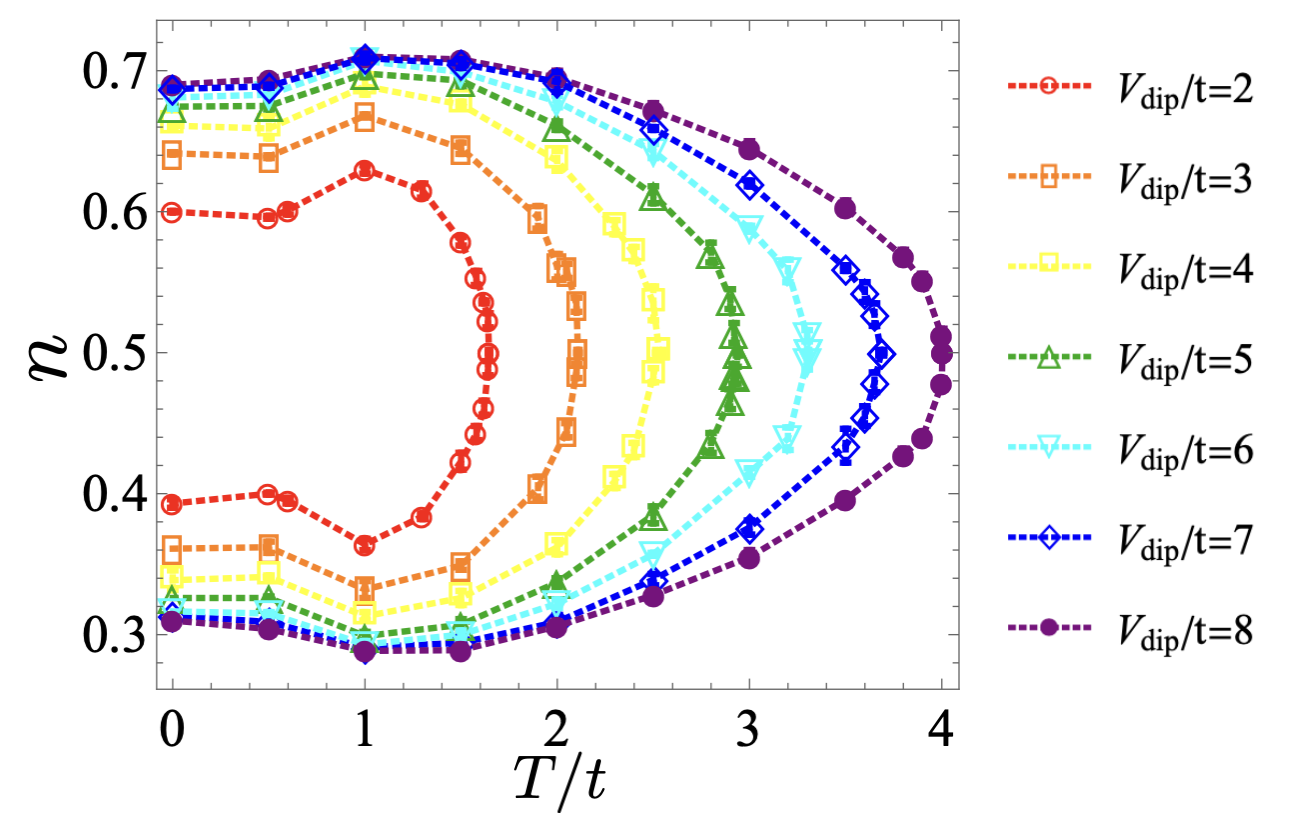} 
    \caption{Transition lines for disappearance of CB solid order at  $V_{\text{dip}}=2$(blue), 3(yellow), 4(green), 5(purple), 6(black), 7(red), 
    8(magenta). All curves exhibit a re-entrant behavior. Transition points have been determined using standard finite-size scaling arguments as discussed in fig.~\ref{fig2}(b) with the difference that here we have scanned through densities at fixed T, rather than the opposite as shown in fig.~\ref{fig2}(b).}
    \label{fig4}
\end{figure}

Finally, in fig.~\ref{fig4} we plot the transition lines for the disappearance of CB density-density correlations for a range of $V_{\text{dip}}/t$ from the onset of CB order to before the onset of a variety of other incompressible phases at $T=0$~\cite{PhysRevA.107.043318}. For each fixed $V_{\text{dip}}/t$, CB order exists inside the corresponding lobe. We notice that the re-entrant behavior is always observed, though it is more pronounced at lower values of $V_{\text{dip}}/t$ where cavity-mediated interactions are more prominent. As expected, at larger values of $V_{\text{dip}}/t$, CB order survives at larger temperatures. These results confirm that at finite $T$, CB order still exists for filling factor as low as $n\sim 0.29$. Therefore, this solid order has the potential to be observed with ultracold polar molecules for which currently achievable filling factors are still pretty low, $n\sim0.25-0.3$~\cite{science.aac6400,PhysRevLett.118.073201}.

\section{Conclusion}
\label{sec:sec5}
In this work, we investigated the effects of finite temperature on a system of  lattice dipolar bosons coupled to a high-finesse optical cavity. We found that there exists a range of densities for which the system undergoes thermo-crystallization above the KT-transition. In other words,  upon increasing temperature, a SF state first becomes a normal liquid and then develops density-density correlations. This effect is due to finite cavity-mediated interactions which suppress exchanges of identical particles hence favoring the solid order. 
We also observe that solid checkerboard order exists at finite temperature for filling factors as low as $n\sim 0.29$ which is close to what is currently achievable in experiments with cold polar molecules. Finally, we found that temperatures needed for observation of off-diagonal and diagonal order are about half hopping amplitude and a few hopping amplitudes respectively. 

\vspace*{12pt}
\begin{acknowledgments}
C. Zhang acknowledges support from the National Natural Science Foundation of China (NSFC) under Grant No. 12204173 and No. 12275263, and the Innovation Program for Quantum Science and Technology (under Grant No. 2021ZD0301900). M. B. acknowledges support from the Natural Sciences and Engineering Research Council of Canada. The computing for this project was performed at the cluster at Clark University.
\end{acknowledgments}

\bibliography{cavity+dipolar} 

\end{document}